%% file: main.tex
\documentclass[letterpaper, 10 pt, conference]{ieeeconf}  % Comment this line out
\IEEEoverridecommandlockouts                              % This command is only
\usepackage{graphics} % for pdf, bitmapped graphics files
\usepackage{amsmath} % assumes amsmath package installed
\usepackage{amssymb}  % assumes amsmath package installed
\usepackage{tikz}
\usetikzlibrary{shapes,arrows}
\usetikzlibrary{positioning}
\usetikzlibrary{calc}

\title{\LARGE \bf
The Unscented Transform Controller: a new model predictive control law for highly nonlinear systems
}

\author{Anna Clarke and Per Olof Gutman% <-this % stops a space
\thanks{This work was not supported by any organization}% <-this % stops a space
\thanks{A. Clarke is with Technion Autonomous Systems Program (TASP),
        Technion - Israel Institute of Technology, Haifa 32000, Israel
        {\tt\small anna.clarke@technion.ac.il}}%
\thanks{P.O. Gutman is with the Faculty of Civil and Environmental Engineering, Technion - Israel Institute of Technology, Haifa 32000, Israel
        {\tt\small peo@technion.ac.il}}%
}

\begin{document}

\maketitle
\thispagestyle{empty}
\pagestyle{empty}

%%%%%%%%%%%%%%%%%%%%%%%%%%%%%%%%%%%%%%%%%%%%%%%%%%%%%%%%%%%%%%%%%%%%%%%%%%%%%%%%
\begin{abstract}
The Unscented Transform which is the basis of the Unscented Kalman Filter, UKF, is used here to develop a novel predictive controller for non-linear plants, called the Unscented Transform Controller, UTC. The UTC can be seen as the dual of the UKF, the same way as the LQG regulator and the Kalman Filter are related. The UTC is demonstrated on the control of complex maneuvers in free fall of a virtual skydiver the model of which was verified in wind tunnel and free fall experiments.

\end{abstract}

%%%%%%%%%%%%%%%%%%%%%%%%%%%%%%%%%%%%%%%%%%%%%%%%%%%%%%%%%%%%%%%%%%%%%%%%%%%%%%%%
\section{INTRODUCTION}\label{sec1}
Even after more than one century of research, non-linear control is still more of an art than an orderly engineering discipline. A plethora of methods and approaches exist, \cite{c5}, from various linearization approaches to more recent neural network controllers, and non-linear model predictive control, \cite{c2}. As a general rule, the control designer must profoundly understand the controlled plant and its particular non-linear structure in order to succeed in choosing an appropriate control method.

In a similar way, non-linear filtering and estimation, \cite{c7}, is an art in contrast to the systematic way linear estimation is done with the Kalman filter, \cite{c9}. For non-linear processes, developments of the Kalman filter include the Extended Kalman Filter (EKF), particle filters, and in particular the numerically efficient Unscented Kalman Filter, UKF, \cite{c6,c8}.

A well known feature of linear control and estimation is the duality of the LQG controller and the Kalman filter, \cite{c10}. In this paper, a novel controller dual of the UKF, called the Unscented Transform Controller, UTC, is proposed for the control of nonlinear plants. The UTC is model based, and can be seen as an original model predictive controller. 

The inspiration behind the UTC is the successful use of the UKF for the estimation of inputs to a highly non-linear skydiver model in complex free-fall maneuvers, and the attempt to control that system. While simple maneuvers for such a model can be efficiently followed using robust linear control, \cite{c11}, such an approach fails for complex maneuvers. Another less than successful attempt was the use of a Recurrent Neural Network Controller, as reported in \cite{c3}. The UTC, however, accurately tracks the desired maneuvers, as demonstrated in this paper, and in \cite{c3} for additional maneuvers such as front and back layouts: transitions between belly-to-earth and back-to-earth equilibria through the head-to-earth orientation.

The paper is organized as follows: Section~\ref{sec2} describes the skydiver model, the control input, and the control task. In Section~\ref{sec3} the Unscented Transform Controller is presented in detail together with some illuminating simulations. Section~\ref{sec4} contains the conclusion with suggestions for future work, and the remark that \cite{c3} contains a rudimentary stability proof for the case the UTC is used to control a linear plant.

\section{SYSTEM UNDER INVESTIGATION}\label{sec2}
The system under investigation is a skydiver performing aerial maneuvers in free-fall. A model of skydiver's body, aerodynamic forces and moments acting on it at terminal velocity (around 60 (m/s)), and equations of motion were developed in \cite{c1}. The model was validated in free-fall and wind tunnel experiments with different skydivers performing a variety of maneuvers. The model is driven by the following inputs: 1. body posture; 2. roll, pitch, and yaw damping moment coefficients; and 3. input moment coefficients, associated with body segments. The damping moment coefficients reflect  an  approximate  body  resistance  (e.g.   muscle stiffness) to the developing rotation rates. The input moment coefficients reflect the consciously applied physical resistance of each limb to the aerodynamic force acting on it. The body posture is defined by the relative orientation of body limbs, while each joint has 3 rotational degrees-of-freedom (DOFs). During skilled actions the human central nervous system organizes these DOFs into \textit{movement patterns}: combinations of DOFs that are activated synchronously and proportionally, as a single unit. These movement patterns can be extracted from free-fall experiments by  recording a sequence of measured postures during the maneuver time and conducting Principal Component Analysis (PCA). It was discovered that experienced skydivers  can perform most maneuvers by utilizing just one movement pattern \cite{c4}. These maneuvers can be reconstructed in simulation by actuating the skydiver model with the chosen movement pattern, the amplitude of which is the control variable. 

For example, in \cite{c3} the aerial rotations maneuver (360 degrees right and left turns in a belly-to-earth position) is reconstructed in simulation by applying a proportional controller with a feed-forward part, as shown in \eqref{eq:lim}.
\begin{equation}
\begin{gathered}
u(t) =1.5 \cdot (yr_{ref}(t)-yr_{real}(t))+0.1\cdot yr_{ref}(t) \\
    -1.5 \leq u(t) \leq 1.5 (rad), \quad -3.5 \leq \frac{du(t)}{dt} \leq 3.5 (rad/s) \\
    pose(t) = NeutralPose + u(t) \cdot TurningPattern
    \label{eq:lim}
    \end{gathered}
\end{equation}
where $yr_{ref}(t)$ $(rad/s)$ is the reference yaw rate profile; $yr_{real}(t)$ $(rad/s)$ is the yaw rate of the virtual skydiver in simulation; $u(t)$ $(rad)$ is the turning pattern angle command; $NeutralPose$ $(rad)$ is the trimmed pose of a skydiver falling straight down, defined by a vector containing values of body DOFs for all modeled joints; and $TurningPattern$ is an eigenvector defining the movement pattern utilized for turning. 

If the reference signal has a slow dynamics, the damping moment coefficients may remain constant during the maneuver, and zero input moments can be assumed. However, as the amplitude and/or the frequency of the desired yaw rate increases it becomes impossible to track the reference signal just by  means of posture adjustments. The reason is that the magnitude of the movement pattern is limited due to the natural constraints of the human body. Additionally, the range of orientations of each limb relative to the airflow, within which the posture preserves aerodynamic efficiency, is also limited. Therefore, the adjustment of input and damping moments becomes necessary. Consider, for example, the yaw rate reference signal in \eqref{eq:yr_com}.
\begin{equation}
    yr_{ref}(t)=400\cdot \frac{\pi}{180}sin(2\cdot \pi \cdot 0.2 \cdot t)
    \label{eq:yr_com}
\end{equation}
\begin{figure}[htb]
%\hspace{-1.5cm}
\centering
\includegraphics[width=0.4\textwidth]{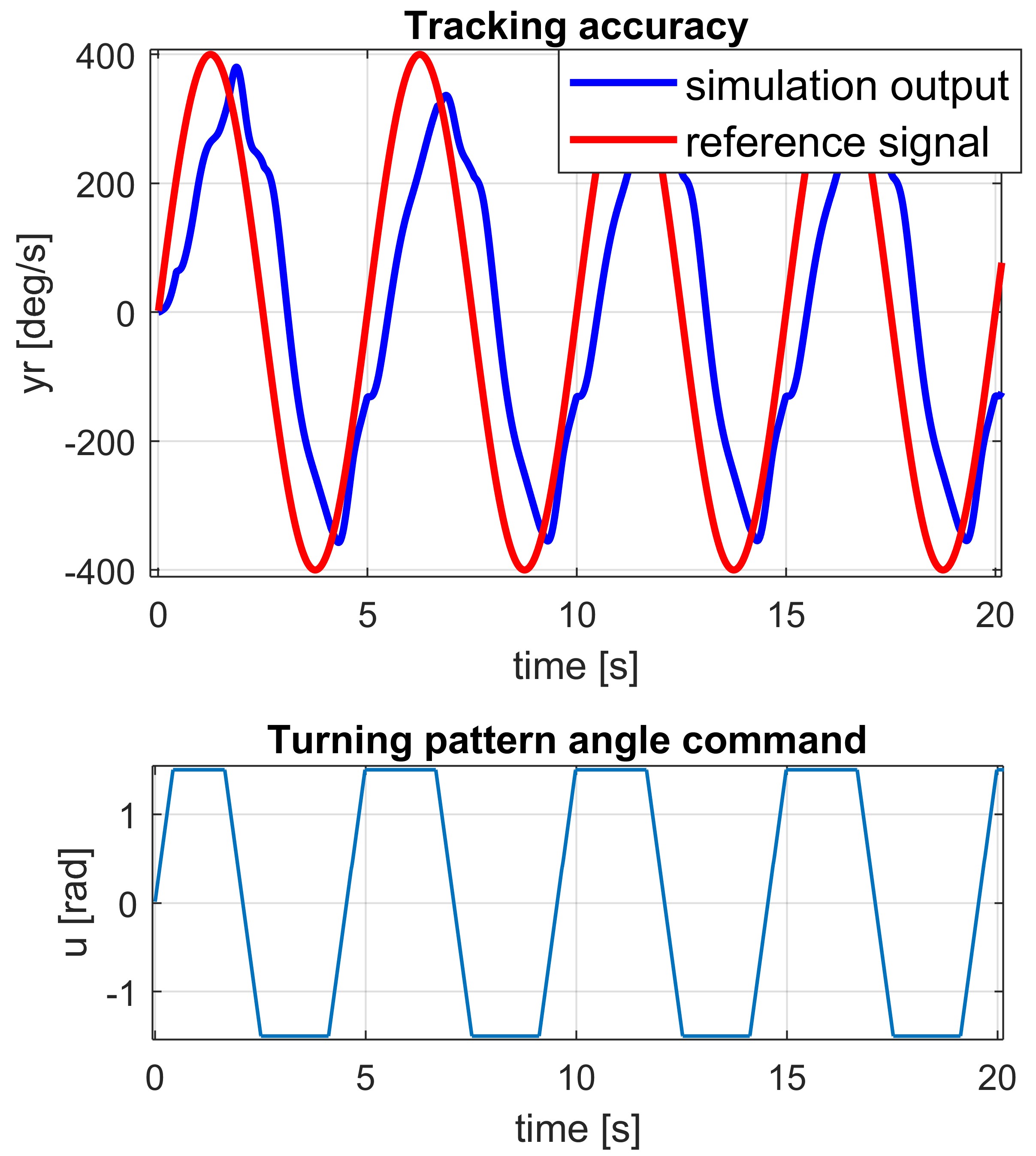}    
\caption{Tracking the yaw rate reference profile with constant yaw damping moment coefficient (0.5) and pose controller in \eqref{eq:lim}.}
\label{fig:ff6}
\end{figure}
This signal can not be accurately tracked by utilizing only a turning movement pattern, while the yaw damping moment coefficient remains constant, see Fig. \ref{fig:ff6}. Thus, for the purpose of an illustrating example of the control method described in the following section, three additional control variables are utilized: the yaw damping moment coefficient, and two input moment coefficients.  

For the purpose of simulating aerial rotations, the following simplified model for the input moments is used:
 \begin{equation}
    \begin{gathered}
        Mz_{hands} = K_{N}\cdot lx_{hand} \cdot (k_{right}-k_{left})  \\
        Fy_{hands} = -K_{N}\cdot (k_{right}+k_{left}) \label{eq_force2}
    \end{gathered}
    \end{equation}
where $Mz$ (Nm) is the roll moment acting about the longitudinal body axis, $Fy$ (N) is the force acting along the sagittal axis, $lx_{hand}$=0.35 (m) is the characteristic (for neutral pose) distance of hands from the longitudinal axis projected to the frontal plane, $K_{N}$=100 (N) is a scaling coefficient, and $k_{right}$,  $k_{left}$ are the user input dimensionless coefficients related to the right and left hands, accordingly. They represent pressure applied  on the airflow with the hands. This creates a roll moment, which will cause the skydiver to develop a roll angle and therefore partially expose his torso to the airflow during the turn. This will create a force along a frontal axis, which will create a yaw moment, desired for the turning maneuver.

\section{UNSCENTED TRANSFORM CONTROLLER}\label{sec3}

This control law was inspired by a modification to the Unscented Kalman Filter developed in  \cite{c1} for estimation of user inputs from experiments of advanced skydiving maneuvers in free-fall. At the prediction step the sigma points and the skydiver state were propagated during the prediction horizon $t_{pred}=0.25$~(s). Allowing each sigma point to drive the skydiver dynamics during the prediction window was essential in order to determine what influence a certain combination of user inputs has on the skydiver state. The prediction time reflected the skydiver model time constant.

Notice, that estimating the user inputs that explain the measured plant dynamics is a very similar problem to predicting the user inputs that will provide the desired maneuvers. The latter is basically the definition of the control problem under investigation: designing a control algorithm, which will compute commands for body posture, input moment coefficients, and damping moment coefficients in order to track the reference angular/linear velocity signals.

Therefore, we suggest the novel control scheme outlined below, which was termed the Unscented Transform Controller (UTC). It has much in common with the concept of standard Model Predictive Control, MPC, \cite{c12}. However it has some important differences: 
\begin{itemize}
    \item The control output is not a result of optimization of a cost function, as in MPC, but a weighted average of sigma points. This average also represents an optimal solution in the sense of Minimum Mean Square Error (MMSE), given the sigma points. 
    \item Propagation of the plant for 4 control variables (i.e. 9 sigma points) is less computationally intensive than solving a nonlinear optimization problem at each step for the same prediction horizon. Additionally, there are no convergence problems that optimization engines have to deal with.
    \item The control outputs do not need to be parameterized as
    in non-linear MPC \cite{c2}, what also contributes to the simplicity of the problem formulation, tuning, and computations.
    \item It is possible to incorporate the desired/expected dynamics of the control signals into the prediction step, when the sigma points are propagated. For example, it can be enforced that the turning pattern angle command has the dynamics of Proportional-Integral (PI) control. 
\end{itemize}

These features are demonstrated subsequently.

\subsection{Formulation}
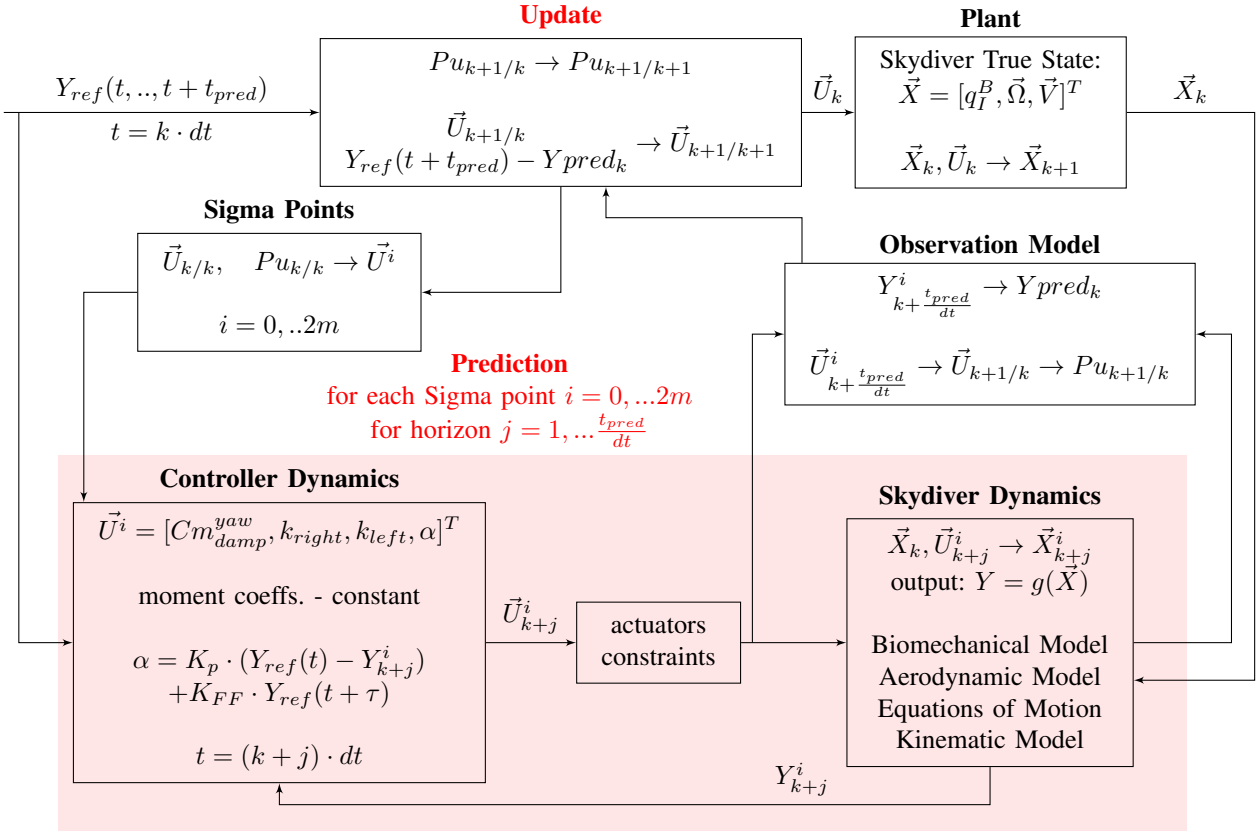
\begin{figure*}[!t]
%\hspace{-1.5cm}
\input{utc_diagram} 
\caption{Block diagram of the Unscented Transform Controller, comprising the computation of sigma points, prediction step, observation model, and update step.} 
\label{fig:utc_diag}
\end{figure*}

The structure of this controller is shown in block diagram in Fig. \ref{fig:utc_diag}: it includes the definition of initial conditions, prediction step, propagation of the plant model up to the prediction horizon, and update step.

%\vspace{0.8cm}
\subsubsection{Control Variables}
%\vspace{0.25cm}

The control variables include the input moment coefficients, damping moment coefficients, and angles of the involved movement patterns. Specifically, in the case of rotation maneuvers:
\begin{equation}
    \vec{U}_k=[Cm_{damp}^{yaw}, k_{right}, k_{left}, \alpha]^T \in \textbf{R}^m,\; m=4
\end{equation}
where $k$ is the simulation step such that $t=k \cdot dt$; $k_{right}, k_{left}$ are the input moment coefficients, see \eqref{eq_force2}; $Cm_{damp}^{yaw}$ is the yaw damping moment coefficient; and $\alpha$ (rad) is the angle of the turning movement pattern, so that the body posture is then computed as:
\begin{equation}
    pose_k=NeutralPose+\alpha_k \cdot TurningPattern
\end{equation}
 If no a-priori knowledge is assumed about the dynamics of these control variables, the controller dynamics becomes:
 \begin{equation}
     \vec{U}_{k+1}=\vec{U}_k+\vec{w_u}_k,  \qquad \vec{w_u}_k \sim N(0,Qu)
 \end{equation}
 where $\vec{w_u}_k$ is the process noise, which will be discussed later, as it is the primary tuning parameter used for obtaining different solutions to the tracking problem.
%\vspace{0.5cm}

\subsubsection{Reference Signal}
%\vspace{0.25cm}

The desired angular/linear velocities play in UTC the same role as the measurement vector in a conventional UKF. In this section it is the yaw rate reference signal (dimension $l=1$), which has to be tracked, defined in \eqref{eq:yr_com}. Thus,
\begin{equation} \label{eq:meas_utc}
    Y_{ref}(k \cdot dt + t_{pred})=yr_{ref}(k \cdot dt + t_{pred}) 
\end{equation}
where $dt$ is  the time between step $k$ and $k+1$, and $t_{pred}$ is the prediction horizon. The measurement noise covariance matrix present in UKF, is denoted in UTC by $P_{err}$ and defines how accurately the reference signal will be tracked, i.e. the trade-off between  accuracy and control effort. 
%\vspace{0.5cm}

\subsubsection{Initial Conditions}
%\vspace{0.25cm}

The initial conditions (step $k=0$) are chosen as follows:
\begin{equation}
    \begin{gathered}
      \vec{U}_{k/k}=[0.5, 0, 0, 0]^T \\
      Pu_{k/k}=\begin{bmatrix}
6.25 & s_{12}\cdot0.74 & s_{13}\cdot0.74 & s_{14}\cdot0.56 \\ s_{12}\cdot0.74 & 0.01 & 0 & 0 \\ s_{13}\cdot0.74 & 0 & 0.01 & 0 \\ s_{14}\cdot0.56 & 0 & 0 & 0.01
\end{bmatrix}
    \end{gathered}
\end{equation}
where $s_{12}, s_{13}, s_{14}$ are the signs of the relevant covariance elements. There are several options for defining $s_{12}, s_{13}$ as a function of $k$, which will be discussed later. The initial $s_{14}=-1$, meaning that in order to start turning the yaw damping moment must be reduced. Equation \eqref{eq:s14} below defines $s_{14}$ as a function of $k$:
\begin{equation}
\begin{gathered}
\Delta(k)=|yr_{ref}(k+1)|-|yr_{ref}(k)| \\
\Delta_{hyst}(k)=|yr_{ref}(k+k_{hyst}+1)|-|yr_{ref}(k+k_{hyst})|\\
s_{14}(k) = \begin{cases}
-sign(\Delta(k)), & \quad s_{14}(k-1)>0 \\
-sign(\Delta_{hyst}(k)), & \quad s_{14}(k-1) \le 0 
\end{cases}
\end{gathered}
    \label{eq:s14}
\end{equation}
where $k_{hyst}$ is set around $\frac{1 (s)}{dt}$ and it is used to provide some hysteresis: the yaw damping moment is increased about 1 second before the turning rate command starts slowing down. Additionally, the following parameters are chosen:
\begin{equation}
    \begin{gathered}
    P_{err}=0.01^2 \\
    Qu=Pu_{k=0} \cdot 0.1
    \end{gathered}
\end{equation}

%\vspace{0.25cm}

\subsubsection{Sigma Points}
%\vspace{0.25cm}

At each simulation step we choose $2m+1$ sigma points $\vec{U^i}$ and their associated weights $W^i$ in the following way, where the value of $W^0$ was chosen by tuning:
\begin{equation} 
    \begin{gathered}
      \vec{U^0}=\vec{U}_{k/k} \\
      \vec{U^i}=\vec{U}_{k/k} \pm \sqrt{\frac{n}{1-W^0}}\vec{S}_j, \qquad  i=1,..2m, \quad j=1,..m \\
      W^i=\frac{1-W^0}{2m}, \qquad W^0=0.25
    \end{gathered}
    \label{eq:sigp}
\end{equation}
where $\vec{S}_j$ is column $j$ of matrix $S$ that satisfies $S \cdot S = Pu_{k/k}$.

\subsubsection{Actuation Constraints}
%\vspace{0.25cm}

After the sigma points are selected they are bounded by the minimum and maximum values of the actuation constraints:
\begin{equation} \label{eq:minmax_utc}
    \begin{gathered}
\vec{U^i}_{min}=[0.01, 0, 0, -1.5]^T \\
\vec{U^i}_{max}=[6, 5, 5, 1.5]^T
 \end{gathered}
\end{equation}
Additionally, the turning pattern angle command $\alpha$ has a slew rate limit, according to \eqref{eq:lim}. This is taken into account when the plant model is propagated: the turning pattern angle at each step is computed from $\vec{U^i}$ and the slew rate constraint.

The sigma points are prevented from being identical, what can happen after imposing the constraints, by modifying the relevant elements of sigma points. If an element $j$ of $ \vec{U}_{k/k}$ equals to its min or max boundary and $\vec{U^i}(j)<\vec{U^i}_{min}(j)$ or $\vec{U^i}(j)>\vec{U^i}_{max}(j)$, respectively, then this element is moved to the other side of the boundary:
\begin{equation} \label{eq:movep}
            \vec{U^i}(j)=1.5 \cdot \vec{U}_{k/k}(j)-0.5 \cdot \vec{U^i}(j)
\end{equation}
%\vspace{0.2cm}

\subsubsection{Observation Model: Propagation of the plant dynamics}
%\vspace{0.25cm}

Each sigma point drives a skydiver model, since the control variables contained in $\vec{U^i}$ are the plant actuators. The plant equations are propagated from the plant's state at step $k$ ($\vec{X_k}$) until the prediction horizon ($t_{pred}$). The plant's state contains the linear and angular velocities of the skydiver. The final value of yaw rate $Y^i_{k+\frac{t_{pred}}{dt}}$, computed for each sigma point $i$ and saved as $Y^i$, is then weighted, \eqref{eq:zpred}, and used for the update step.   
\begin{equation}
          \begin{gathered}
          Y^i \text{ is } Y^i_{k+\frac{t_{pred}}{dt}}\\ 
          Ypred_{k}= \displaystyle \sum_{i=0}^{2m}W^iY^i \\  W^i=\frac{1-W^0}{2m}
          \end{gathered}
          \label{eq:zpred}
         \end{equation}
The prediction horizon $t_{pred}$ is a tuning parameter and will be discussed later in this section.
%\vspace{0.5cm}       

\subsubsection{Prediction Step}
%\vspace{0.25cm}

        The computation flow is schematically shown in Figure \ref{fig:utc_diag}. Sigma points are selected according to  (\ref{eq:sigp}), constrained according to  (\ref{eq:minmax_utc}), (\ref{eq:movep}), propagated until the prediction horizon $\vec{U^i}_{k+\frac{t_{pred}}{dt}}$, saved as $\vec{U^i}$,  and summarized as:
        \begin{equation}
        \begin{gathered}
        \vec{U^i} \text{ is } \vec{U^i}_{k+\frac{t_{pred}}{dt}} \\ 
        \vec{U}_{k+1/k}=\displaystyle \sum_{i=0}^{2m}W^i\vec{U^i} 
        \end{gathered}
          \label{eq:xpred}
         \end{equation}
         
         \begin{equation}
          Pu_{k+1/k}=Qu+\displaystyle \sum_{i=0}^{2m}W^i(\vec{U^i}-\vec{U}_{k+1/k})\cdot(\vec{U^i}-\vec{U}_{k+1/k})^T
          \label{eq:pu_pred}
         \end{equation}

    \subsubsection{Update Step}
    
        \begin{equation}
            C_y=P_{err}+\displaystyle \sum_{i=0}^{2m}W^i(Y^i-Ypred_{k})\cdot(Y^i-Ypred_{k})^T
        \end{equation}
         \begin{equation}
            C_{uy}=\displaystyle \sum_{i=0}^{2m}W^i(\vec{U^i}-\vec{U}_{k+1/k})\cdot(Y^i-Ypred_{k})^T
        \end{equation}
        \begin{equation}
            \begin{gathered}
            K=C_{uy}C_y^{-1} \\
            \vec{U}_{k+1/k+1}=\vec{U}_{k+1/k}+K(Y_{ref}(k\cdot dt+t_{pred})-Ypred_{k}) \\
            Pu_{k+1/k+1}=Pu_{k+1/k}-KC_yK^T 
            \end{gathered}
            \label{eq:utc_controller}
        \end{equation}
        The updated controller $\vec{U}_{k+1/k+1}$ is bounded according to  (\ref{eq:minmax_utc}), and the slew rate constraint  (\ref{eq:lim}).
   
%\vspace{0.5cm}

\subsection{Simulation Results}

The fast turning maneuver defined in Sect. \ref{sec2} is now tracked utilizing the UTC, while different solutions are obtained by the means of configuring the controller parameters.

First of all, the input moment coefficients $k_{right}$, $k_{left}$ are considered. Their desired dynamics is not known and, therefore, the behavior is driven by the process noise $Qu$. Thus, in order to get qualitatively different solutions the off-diagonal values of $Qu$ are modeled in different ways, e.g. \eqref{eq_q1}, \eqref{eq_q3}.
\begin{equation}
    \begin{gathered}
    Qu_{1,2} = Qu_{2,1} = q_{1,2} \cdot sign(yr_{ref}) \\
    Qu_{1,3} = Qu_{3,1} = -q_{1,3} \cdot sign(yr_{ref})
    \end{gathered}
    \label{eq_q1}
\end{equation}
where $1,2,3$ are the indices of $Cm_{damp}^{yaw}$, $k_{right}$, $k_{left}$,  respectively, and $q_{1,2}, q_{1,3}$ are tuning parameters that can be assumed equal.
\begin{equation}
    \begin{gathered}
    Qu_{1,2} = Qu_{2,1} = q_{1,2} \cdot sign(\frac{d}{dt} yr_{ref}) \\
    Qu_{1,3} = Qu_{3,1} = q_{1,3} \cdot sign(\frac{d}{dt}yr_{ref})
    \end{gathered}
    \label{eq_q3}
\end{equation}
The coupling in \eqref{eq_q1} means that when $Cm_{damp}^{yaw}$ is reducing, making the turn faster, the appropriate input moment is increased to further accelerate the turn. Thus, the coupling expressed by $Qu$ should be negative for $ k_{left}$ during left turns ($sign(yr_{ref})>0$) and for $ k_{right}$ during right turns, and positive otherwise. 

The desired dynamics of the turning pattern angle command ($\alpha$) can be assumed unknown, and left to be resolved by the UTC. Optionally, the dynamics given in \eqref{eq:lim} can be utilized. The gains of the proportional and the feed-forward parts can be tuning parameters. The computation of the controller command during the prediction window can be implemented as shown in \eqref{eq:m1utc}. This enforces the slew rate constraint on $\alpha(t)$, see \eqref{eq:lim}.
\begin{equation}
        \begin{split}
            \alpha(t)= \alpha(t-dt)+ & K_p \cdot (yr_{ref}(t)-yr_{ref}(t-dt))- \\ -&K_p \cdot (yr_{real}(t)-yr_{real}(t-dt))+ \\ +&K_{FF} \cdot (yr_{ref}(t+\tau)-yr_{ref}(t+\tau-dt))
            \end{split}
            \label{eq:m1utc}
        \end{equation}

\begin{figure}[!htb]
%\hspace{-1.5cm}
\includegraphics[width=0.5\textwidth]{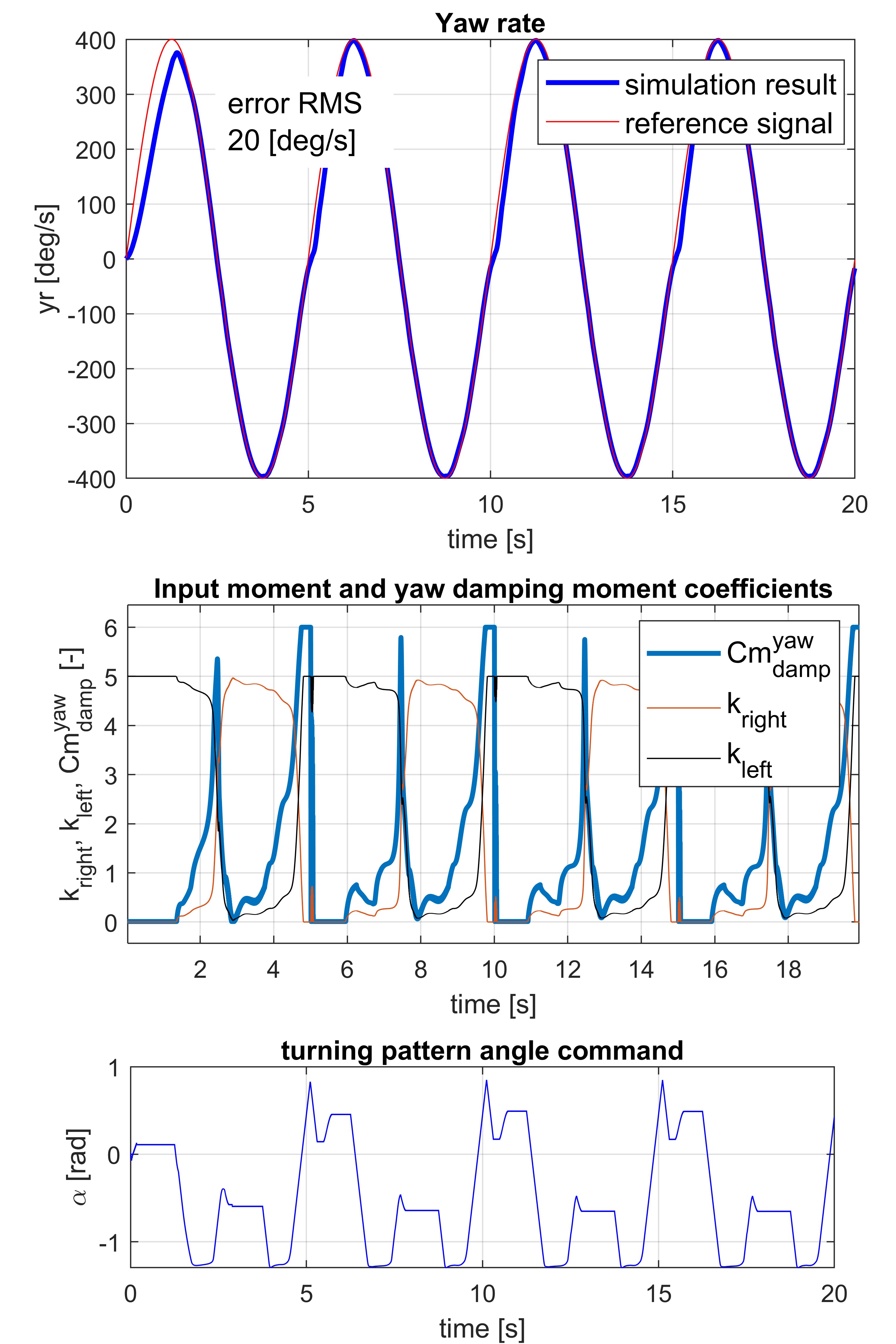}  
\caption{Tracking the yaw rate reference profile with three user input coefficients and body posture controlled by the Unscented Transform Controller, assuming unknown controller dynamics and correlation between input moment and yaw damping moment coefficients according to  \eqref{eq_q1}.} 
\label{fig:utc_m0_c2}
\end{figure}
\begin{figure}[!htb]
%\hspace{-1.5cm}
\includegraphics[width=0.5\textwidth]{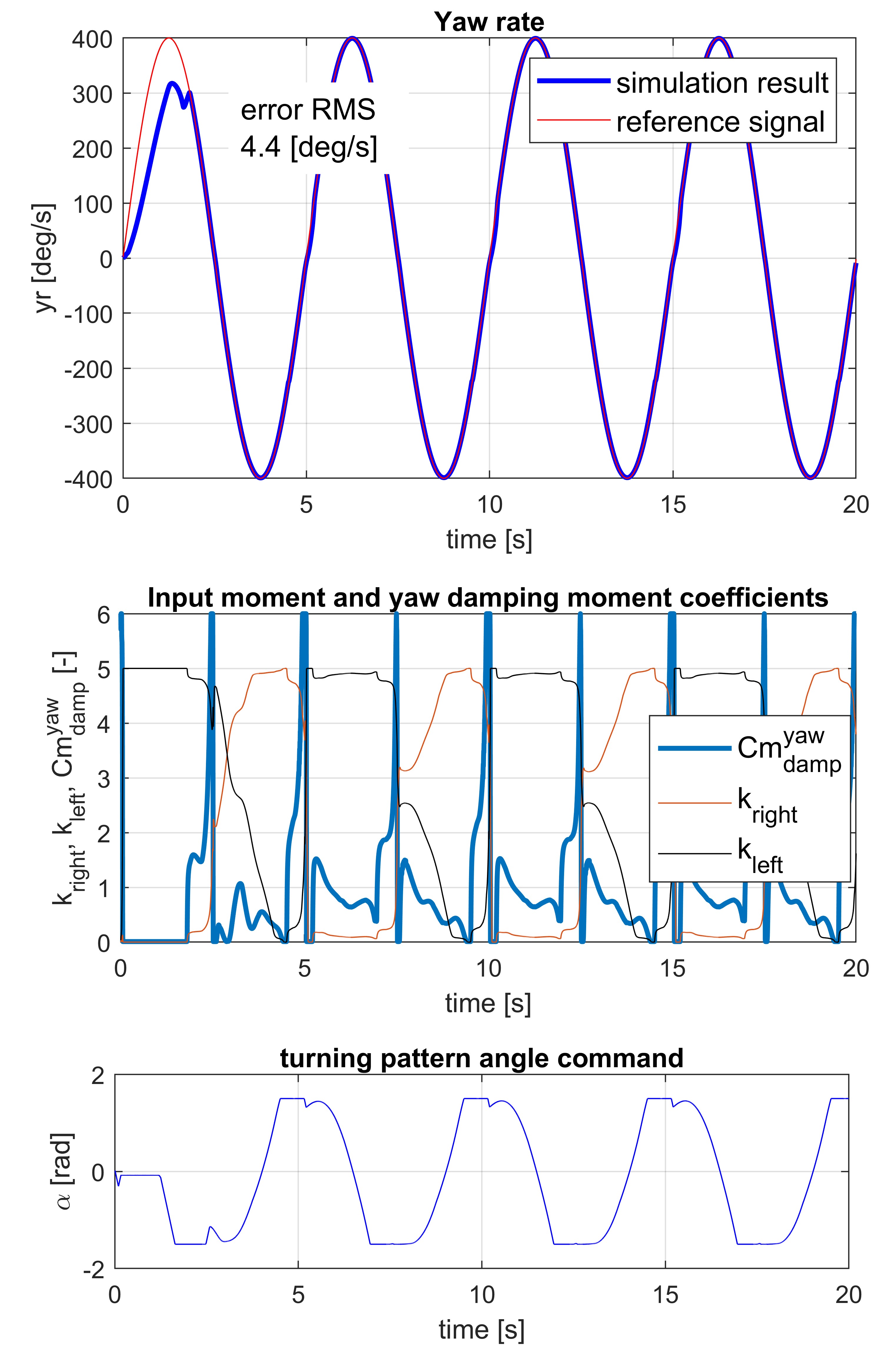}  
\caption{Tracking the yaw rate reference profile with three user input coefficients and body posture controlled by the Unscented Transform Controller, assuming controller dynamics according to \eqref{eq:m1utc} and correlation between input moment and yaw damping moment coefficients according to \eqref{eq_q1}} 
\label{fig:utc_m1_c2}
\end{figure}
\begin{figure}[!htb]
%\hspace{-1.5cm}
\includegraphics[width=0.5\textwidth]{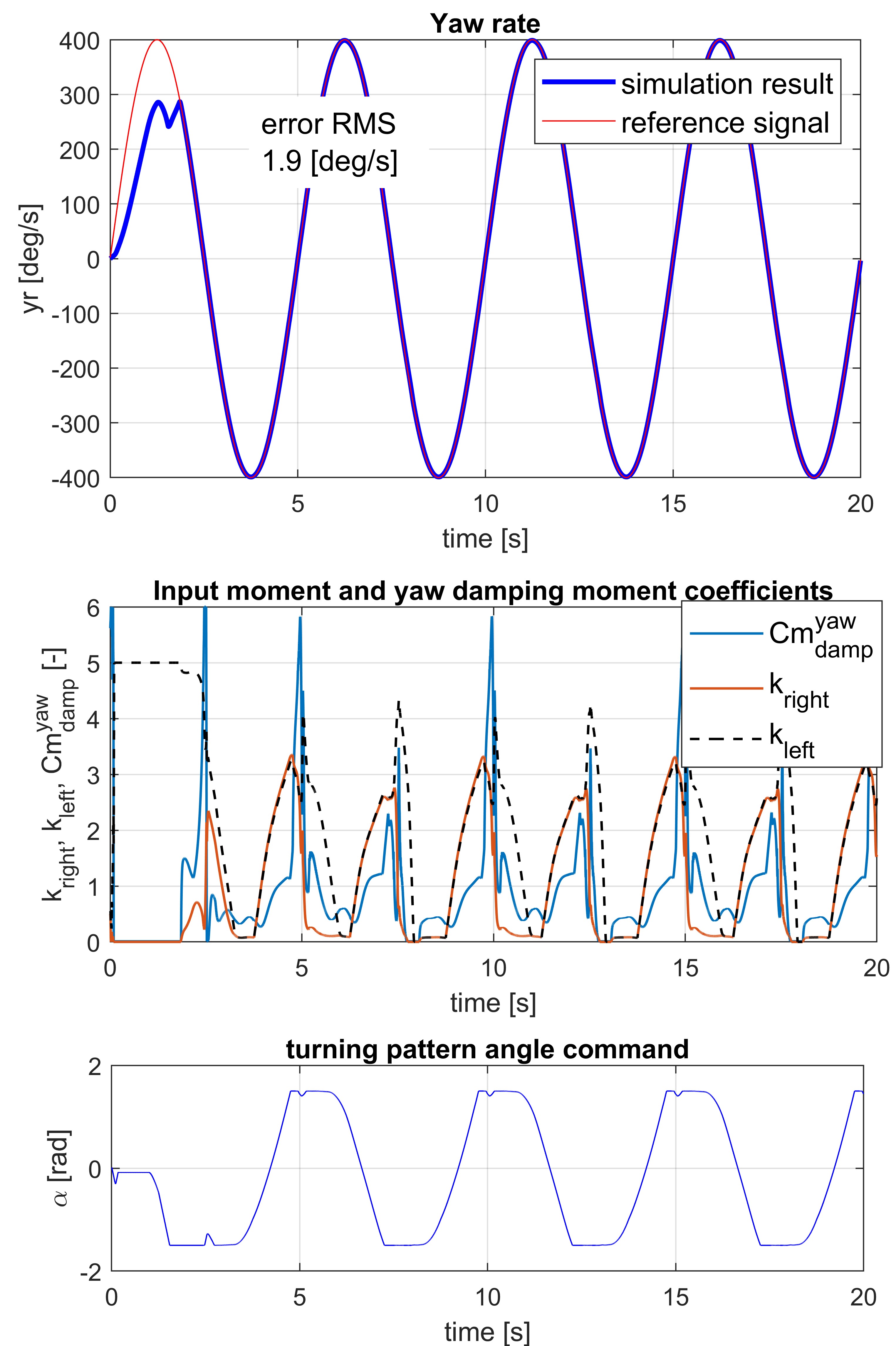}  
\caption{Tracking the yaw rate reference profile with three user input coefficients and body posture controlled by the Unscented Transform Controller, assuming controller dynamics according to  \eqref{eq:m1utc} and correlation between input moment and yaw damping moment coefficients according to \eqref{eq_q3}} 
\label{fig:utc_m1_c3}
\end{figure}

 When a desired dynamics for the turning pattern angle is incorporated in the prediction step, a better accuracy is achieved: compare Figures \ref{fig:utc_m0_c2}, \ref{fig:utc_m1_c2}. Also a smaller prediction horizon can be used. For $dt=0.0042 \, (s)$ 20 prediction steps were used (i.e. $t_{pred}\approx0.08 \, (s)$) when assuming unknown dynamics for $\alpha(t)$, versus 5 steps ($t_{pred}\approx0.02 \, (s)$) when incorporating \eqref{eq:m1utc}. 
   
Notice the qualitatively different behavior of the input moment coefficients, shown in Figs. \ref{fig:utc_m1_c2}, \ref{fig:utc_m1_c3}. 
If $\alpha(t)$ dynamics is unknown, the best accuracy is achieved when the correlation between the yaw damping moment and the input moment coefficients is defined according to  \eqref{eq_q1}, see Fig. \ref{fig:utc_m0_c2}. Notice, that \eqref{eq_q1} is a more intuitive model than \eqref{eq_q3}. If $\alpha(t)$ dynamics is reasonably modeled, tuning the correlation between these variables is less significant. Accurate tracking is achieved even if no correlation is assumed. This, and other interesting solutions, are shown in  \cite{c3}.
The minimal control effort is achieved when combining \eqref{eq:m1utc} and \eqref{eq_q3}, see Fig. \ref{fig:utc_m1_c3}.

\section{CONCLUSIONS AND FUTURE WORK}\label{sec4}
From simulations we observe the following advantages of the UTC: 1. Fast convergence (around 2 (s)); 2.
    Not much sensitivity to the tuning parameters: if the covariances are set to reasonable values, i.e. the same order of magnitude as the allowed ranges of the actuators involved, UTC always converges. 3. Convenient tuning: achieved by the means of setting the elements of $Pu_0$ and $Qu$, which have a clear physical meaning. Therefore, it is possible to obtain different solutions that implement the desired/ expected behavior of the control variables.

Further investigating the UTC properties remains an issue for future work: sensitivity to model uncertainty and measurement noise, and operation with only partial state feedback. Additionally, it is important to formulate and prove conditions for the global/ local stability of systems controlled by the UTC. Initial efforts in this regard are presented in \cite{c3}: for linear systems a stability condition takes the form of an Algebraic Riccati Equation, while for the number of prediction steps $n>1$ it has one additional term.
%\addtolength{\textheight}{-3cm}   % This command serves to balance the column lengths

\end{document}

%% file: utc_diagram.tex
\tikzstyle{blockp} = [fill, fill=red!10, rectangle, minimum height=5.0cm, minimum width=15.0cm]
\tikzstyle{blockwW} = [draw, rectangle, minimum height=0.8cm, minimum width=1.2cm]
\tikzstyle{blockw} = [draw, rectangle, minimum height=0.1cm, minimum width=0.2cm]
\tikzstyle{input} = [coordinate]

 \begin{tikzpicture}[auto, node distance=3.7cm, >=latex']
 \hspace{-0.2cm}
\node [input](in0) {};
\node [blockp, right=0.0cm of in0] (b0) {};
\node [blockw, right=0.2cm of in0] (b1) {{\begin{tabular}{c}   $\vec{U^i}=[Cm_{damp}^{yaw}, k_{right}, k_{left}, \alpha]^T$ \\ \\ moment coeffs. - constant \\ \\ $\alpha=K_p \cdot (Y_{ref}(t)-Y^i_{k+j})$ \\ $+K_{FF} \cdot Y_{ref}(t+\tau)$ \\  \\ $t=(k+j) \cdot dt$ \end{tabular}}};

\node [above] at ($(b1.north)+(3.0,0.6)$) {{ \begin{tabular}{c} \textbf{\textcolor{red}{ Prediction}} \\\textcolor{red}{for each Sigma point $i = 0, ... 2m $} \\ \textcolor{red}{for horizon $j=1,... \frac{t_{pred}}{dt}$}\end{tabular}}};

%\node [input, left=1.7cm of b1](in0_0) {};
%\draw[->] (in0) -- node[above] {} (b1.west);
\node [above] at (b1.north) {{\textbf{Controller Dynamics} }};
\node [blockwW, right=1.2cm of b1] (b2){{\begin{tabular}{c} actuators \\ constraints  \end{tabular}}};
%\node [inside] at (b2.center) {{\begin{tabular}{c} actuators \\ constraints  \end{tabular}}};
\node [blockw, right=1.4cm of b2] (b3){{\begin{tabular}{c} $\vec{X}_{k}, \vec{U}^i_{k+j} \rightarrow \vec{X}^i_{k+j}$ \\output:  $Y=g(\vec{X})$\\ \\Biomechanical  Model \\ Aerodynamic  Model \\ Equations of  Motion  \\ Kinematic Model \end{tabular}}};
\node [above] at (b3.north) {{\textbf{Skydiver Dynamics} }};
\node [input, below=0.7cm of b2](in1) {};
%\draw [->] (in1) -- node[above] {{IC: $X_k$}} ($(b3.west) - (0, 1.1)$);
\draw [->] (b2.east) -- (b3.west);
\draw [->] (b1.east) -- node[above] {{$\vec{U}^i_{k+j}$}} (b2.west);
\path [draw,->] (b3.south) -- ($(b3.south) - (0, 0.5)$) -- node[above] {{$Y^i_{k+j}$}}($(b3.south) - (5, 0.5)$) -| (b1.south);

\node [blockw, above=1.5cm of b3] (b4){{\begin{tabular}{c}  $Y^i_{k+\frac{t_{pred}}{dt}} \rightarrow Ypred_{k}$\\ \\ $\vec{U}^i_{k+\frac{t_{pred}}{dt}} \rightarrow \vec{U}_{k+1/k} \rightarrow  Pu_{k+1/k} $  \end{tabular}}};
\path [draw,->] (b3.east) -- ($(b3.east) + (1.3, 0)$) |- (b4.east);
\node [input, right=0.15cm of b2](in2) {};
\path [draw,->] (in2) -- ($(in2) + (0, 2.1)$)  |- (b4.west);
\node [above] at (b4.north) {{\textbf{Observation Model} }};

\node [blockw, above=2.0cm of b1] (b5){{\begin{tabular}{c} $\vec{U}_{k/k}, \quad Pu_{k/k} \rightarrow \vec{U^i}$
      \\ \\
      $  i=0,..2m$
     \end{tabular}}};
\node [above] at (b5.north) {{\textbf{Sigma Points} }};

\node [input, above=2.0cm of b4](in_b6) {};
\node [blockw, left=2.5cm of in_b6] (b6){{\begin{tabular}{c} $Pu_{k+1/k} \rightarrow  Pu_{k+1/k+1}$
\\  \\$\begin{matrix}\vec{U}_{k+1/k}  \\  Y_{ref}(t + t_{pred})-Ypred_{k} \end{matrix} \rightarrow \vec{U}_{k+1/k+1}$
      
       \end{tabular}}};
\node [above] at (b6.north) {{\textbf{\textcolor{red}{Update}} }};
\path [draw,->] (b6.south) -- ($(b6.south)-(0,0.2)$) |- (b5.east);
\path [draw,->] (b5.west) -- ($(b5.west)-(0.2,0)$) -| ($(b1.north)-(2.6,0)$);
\path [draw,->] ($(b4.north)+(-2.5,0)$) -- ($(b4.north)+(-2.5,0.6)$) -| ($(b6.south)+(0.6,0)$);
\node [input, left=4.2cm of b6](in0_b6) {};
\draw[->] (in0_b6) -- node[above] {{$Y_{ref}(t,..,t+t_{pred})$}} node[below] {$t=k \cdot dt$} (b6.west);
\path [draw,->] ($(in0_b6) + (0.2, 0)$)   |- (b1.west);

\node [blockw, right=0.7cm of b6] (b7){{\begin{tabular}{c} Skydiver True State: \\ $\vec{X}=[q_I^B,\vec{\Omega},\vec{V}]^T$ \\ 
\\ $\vec{X}_{k}, \vec{U}_{k} \rightarrow \vec{X}_{k+1}$
       \end{tabular}}};
\draw [->] (b6.east) -- node[above] {$\vec{U}_k$} (b7.west);
\node [above] at (b7.north) {{\textbf{Plant} }};
\path [draw,->] (b7.east) -- node[above] {$\vec{X}_{k}$} ($(b7.east) + (1.7, 0)$) |- ($(b3.east) - (0.0, 0.5)$);

\end{tikzpicture}